\documentclass[aps,prd,floatfix,nofootinbib,showpacs,superscriptaddress]{revtex4}
\usepackage{graphicx,amsmath,amssymb,bm}
\usepackage{color}
\usepackage[utf8]{inputenc}
\definecolor{purple}{rgb}{0.5,0,0.5}
\definecolor{blue}{rgb}{0.0,0,0.9}
\usepackage[colorlinks=true, pdfstartview=FitV, citecolor= purple, linkcolor = blue,urlcolor=blue]{hyperref}
\usepackage{amssymb}
\usepackage{graphics}
\usepackage{multirow}
\usepackage{epsfig}
\usepackage{rotating}
\usepackage{longtable} %\usepackage[english]{babel}
\usepackage{url}
\usepackage{pdflscape}
\usepackage{cancel}
\pdfoutput=1
\usepackage{csquotes}
\usepackage{placeins}
\usepackage{scalerel}
\usepackage{bbm}

\begin{document}
\title{Five Texture Zeros for Dirac Neutrino Mass Matrices}

\author{Richard H. Benavides}
\email{richardbenavides@itm.edu.co}
\affiliation{ Facultad de Ciencias Exactas y Aplicadas, Instituto Tecnol\'ogico Metropolitano,
Calle 73 No 76 A - 354 , V\'ia el Volador, Medell\'in, Colombia}
 \author{Yithsbey Giraldo}
 \email{yithsbey@gmail.com}
 \affiliation{Departamento de F\'\i sica, Universidad de Nari\~no, A.A. 1175, 
 San Juan de Pasto, Colombia}
\author{Luis Muñoz}
\email{luismunoz@itm.edu.co}
\affiliation{ Facultad de Ciencias Exactas y Aplicadas, Instituto Tecnol\'ogico Metropolitano,
Calle 73 No 76 A - 354 , V\'ia el Volador, Medell\'in, Colombia}
\author{William A. Ponce}
\email{william.ponce@udea.edu.co}
\affiliation{Instituto de F\'isica, Universidad de Antioquia, Calle 70 No.~52-21, Apartado A\'ereo 1226, Medell\'in, Colombia}
\author{Eduardo Rojas}
\email{eduro4000@gmail.com}
\affiliation{Departamento de F\'\i sica, Universidad de Nari\~no, A.A. 1175, 
San Juan de Pasto, Colombia}

\begin{abstract}
In this work we propose new five textures zeros for the mass matrices in the lepton 
sector in order to predict values for the neutrino masses. In our approach we go 
beyond the Standard Model by assuming Dirac masses for the neutrinos,  
feature which allows us to make a theoretical prediction for the lightest neutrino 
mass in the normal ordering.
The textures analyzed have enough free parameters to  adjust the $V_{\text{\tiny PMNS}}$ 
mixing matrix including the CP-violating phase, the neutrino mass squared differences 
$\delta m_{21}^2, \delta m_{31}^2$, and the three charged lepton masses. In order to have reliable 
results two different approaches are used: the first one is based on a least-squares 
analysis to fit the lepton masses and the mixing parameters to their corresponding 
experimental values, for this case the best fit for the lightest neutrino mass is 
$(3.9\pm^{0.6}_{0.8})\times 10^{-3}$~eV.; the second approach is just algebraic, based on the weak basis transformation method,  
in this case the lightest neutrino mass consistent with the experimental 
values and the restrictions coming from the five texture zeros of the 
mass matrices is  equal to $(3.5\pm0.9)\times 10^{-3}$~eV.

\pacs{
14.60.Pq,      %Neutrino mass and mixing
14.60.St,      %Non-standard-model neutrinos, right-handed neutrinos, etc.
14.60.Lm.      %Ordinary neutrinos
}
\end{abstract}

\maketitle

\section{Introduction  \label{intro}}
The Standard Model (SM) of the strong and electroweak interactions~\cite{Donoghue:1992dd} based on 
the local gauge group $SU(3)_c\otimes SU(2)_L\otimes U(1)_Y$, with the unbroken 
$SU(3)_c$ sector confining and the electroweak sector $SU(2)_L\otimes U(1)_Y$ 
spontaneously broken by one scalar complex Higgs doublet down to $U(1)_{em}$, has 
been very successful in explaining several experimental issues. Unanswered questions 
of the model are the explanation of the total number of families present in nature, 
the hierarchy of the charged fermion mass spectrum, the quark and lepton mixing 
angles and the origin of CP violation. Even mayor issues are the existence of dark 
energy and dark matter and the smallest of the neutrino masses and their oscillations.

The extension of the SM with three right handed neutrinos provides with several nice 
features: a)  it allows the introduction of nine additional complex Dirac mass terms in the neutral lepton sector~\cite{Peinado:2019mrn},
b)  it permits the implementation of the seesaw mechanism~\cite{Minkowski:1977sc,GellMann:1980vs,Yanagida:1980xy,Glashow:1979nm,Mohapatra:1979ia,Schechter:1980gr}, and c)  it conduces to 
an analysis with hermitian mass matrices in the lepton sector of the model. Although 
the first two features have been widely explored in the literature, the last one has 
not; it is the purpose of this paper to analyze the mathematical and numerical 
consequences of this last fact.

The texture of the Dirac neutrinos mass matrix has been the subject of several 
recent studies~\cite{Ahuja:2009jj,Liu:2012axa,Verma:2013cza,Ludl:2014axa,Verma:2014lpa,
Fakay:2014nea,Cebola:2015dwa,Gautam:2015kya,Ahuja:2016san,Singh:2018lao,Ahuja:2018fmw};
this, in part, has been motivated by the non-observation of the double beta 
decay~\cite{DellOro:2016tmg}. As it is well known, the existing experiments have 
not been able to determine if neutrinos are  Majorana or Dirac particles; 
however, most work on these topics assumes that neutrinos are Majorana, whereas the 
case for Dirac neutrino masses has not been studied properly.
In this analysis we assume that neutrinos are  Dirac particles~\cite{Peinado:2019mrn}, 
which allows us to use the weak  basis transformation, the polar decomposition theorem, 
and the most recent experimental data to make predictions based on a given texture 
for the lepton mass matrices. It is interesting to determine the consequences of a 
given texture on the prediction of the neutrino masses; indeed, for Majorana masses 
there are several works predicting a mass of a few milli-electron volts for the lightest 
neutrino mass~\cite{Fritzsch:2015gxa,Fritzsch:2015haa,Fritzsch:2016xmb}; which, as 
we will see, has the same order of magnitude as the results presented in this work.

There exist top-bottom motivations to assume Dirac masses for the neutrinos; in 
models derived from string theory the Majorana masses  are suppressed by  selection 
rules related to the underlying symmetries~\cite{Antusch:2005kf,Langacker:2011bi}. 
In these models, Majorana masses can also  be generated by active neutrinos via 
gravitational effects; however, these masses are very small, even compared with 
the small scales in neutrino physics~\cite{Dvali:2016uhn,Barenboim:2019fmj}. 
The Dirac neutrino electromagnetic properties are quite testable too due to the 
magnetic dipole moment which is different from zero  at the quantum 
level~\cite{Broggini:2012df}. The same  is not true for Majorana neutrinos, where 
their electromagnetic properties are reduced because they are their own antiparticles.  
In this way, theoretical models with Dirac neutrinos are a motivation to refine 
the electromagnetic constraints on the neutrino properties. 

In the SM extended with three right-handed neutrinos (SMRHN) and forbidding the 
bare Majorana masses of the right-handed neutral states (obtained just by 
assuming lepton number conservation), the lepton mass terms after the spontaneous 
breaking of the local symmetry are given by
\begin{align}\label{lama}
-\mathcal{L}_D=\bar{\nu}_L M_n\nu_R+ \bar{\ell}_L M_\ell \ell_R+\text{h.c},
\end{align} 
where $\nu_{L,R}= (\nu_{e},\nu_{\mu},\nu_{\tau})_{L,R}^T$
and   $\ell_{L,R}= (e,\mu,\tau)_{L,R}^T$
(the upper $T$ stands for transpose). The matrices  $M_n$ and $M_{\ell}$ in 
(\ref{lama}) are in general $3\times 3$ complex mass matrices.  In the most 
general case they contain  36 free parameters. In the context of the SM extended 
with three right-handed neutrinos, such a large number
of parameters can be drastically cut by making use of the polar theorem of matrix 
algebra,  by which, one can always decompose a general complex matrix as the product 
of an hermitian times a unitary matrix. 
Since in the context of the SM extended with three right-handed neutrinos the 
unitary matrix can be absorbed in a redefinition of the right handed lepton 
components, this immediately brings the number of free parameters from 36 down to 18 
(the other eighteen parameters can be hidden in the right-handed lepton components in 
the context of these models and some of its extensions, but not in its left-right 
symmetric extensions).
So, as far as this model is concerned, we may treat without loss of generality 
$M_n$ and $M_{\ell}$ as two hermitian  mass matrices, with 18 real parameters in 
total, out of which six are phases. Since five of those phases can be absorbed in 
a redefinition of the lepton fields~\cite{Kobayashi:1973fv,Maiani:1975in}, the 
total number of free parameters we may play with in $M_n$ and $M_{\ell}$ are 12 
real parameters and one phase; this last one used to explain the CP violation 
phenomena.

Now, in the context of the SMRHN, it is always 
possible to implement the so called weak basis~(WB) 
transformation~\cite{Branco:1988iq,Branco:1999nb}, which leaves the two 
$3\times 3$ lepton mass matrices Hermitian and does not alter the physics implicit in 
the weak currents. Such a WB transformation is a unitary transformation acting 
simultaneously in the neutral and lepton fermion mass 
matrices~\cite{Branco:1988iq,Branco:1999nb}. That is
\begin{equation}\label{wbss}
 M_n\rightarrow M_n^R=UM_nU^\dagger, \hspace{1cm} M_l\rightarrow M_l^R=UM_nU^\dagger,
\end{equation}
where $U$ is an arbitrary unitary matrix. We say then that the two representations 
$(M_n,M_l)$ and $(M_n^R,M_l^R)$ are equivalent in the sense that they are related 
to the same $V_{\text{PMNS}}$ mixing matrix. This kind of transformation plays an important 
role in the study of the so-called flavor problem.

In references~\cite{Branco:1988iq,Branco:1999nb} it was shown that, 
for hermitian quark mass matrices it is always possible to perform a weak 
basis transformation, in such a way that 
it is always possible to end up with two hermitian  mass matrices with three texture 
zeros which do not have any physical implication.
As we mentioned above we extend the SM to include the right-handed neutrinos and consider them to be Dirac
type particles.  Thus,  the mechanism implemented in the quark sector can be transferred,  with few modifications,
directly to the lepton sector. 
With these three non-physical texture zeros the number of free parameters in $M_n^R$ and 
$M_{\ell}^R$ reduces from twelve to nine real parameters and one phase, just enough 
to fit the 
physical values for the six Dirac lepton masses, the three mixing angles, and the 
CP violation phase. Any extra texture zero can only be a physical assumption and 
should imply a relationship between the ten physical parameters.
But for the case of neutrinos, their masses are not known, and just two square mass 
differences are experimentally available; that is: $\delta m_{21}= m_2^2-m_1^2$, 
$\delta m_{31}= m_3^2-m_1^2$. Therefore, only two experimental restrictions 
are present in the analysis of the neutral lepton sector (which is not the case in the quark 
sector), and a fourth texture zero, 
if possible, will not produce a testable prediction at all. The next step, a fifth 
texture zero should provide us with a physical prediction and a possible sixth 
texture zero will provide with two. (Counting degrees of freedom we find that the 
maximum number of such texture zeros 
consistent with the absence of a zero mass eigenvalue and a nondegenerate mass 
spectrum in the lepton sector is just six, with three in the neutral sector 
and three in the charged one.)

In the analysis which follows we are going to use the SM ingredients including one 
single complex Higgs doublet, enlarged with three right-handed neutrinos (one 
for each family) in order to provide the neutral sector with Dirac masses, and an 
unknown symmetry able to produce five texture zeros in the lepton sector. Two cases 
are going to be considered: one with three texture zeros in the neutral lepton 
sector and two in the charged one, and a second case with two texture zeros in the 
neutral lepton sector and three in the charged one. Two different analysis are 
implemented, one for each situation. Analytic and numeric results are presented, 
taking special care to accommodate the latest experimental data 
available~\cite{Tanabashi:2018oca}, including the CP violation phenomena.

%%%%%%%%%%%%%%%%%%%%%%%%%%%%%%%%%%%%%%%%%%%%%%%%%%%%%
\section{Five texture zeros. First case.}
In the context of the SMRHN with lepton number conservation, in the weak basis, 
and after breaking the local gauge symmetry, the Lagrangian mass term for the 
lepton sector is given by
\begin{align}\label{eq:dmass}
-\mathcal{L}_D=\bar{\nu}_L^{\prime} M'_n\nu_R^{\prime}+\bar{\nu}_R^{\prime} 
M'^{\dagger}_n\nu_L^{\prime}+
 \bar{\ell}_L^{\prime} M'_\ell \ell_R^{\prime}+\bar{\ell}_R^{\prime} 
 M'^{\dagger}_\ell \ell_L^{\prime},
\end{align} 
where $M'_n$ and $M_\ell'$ are the neutrino and charged lepton  mass matrices 
respectively (primed  fields and  matrices refer to the weak basis).

Let us now assume that for a given symmetry, the Hermitian mass matrices $M'_n$ 
and $M_\ell'$ present the following textures
\begin{align}\label{eq:un}
 M'_n=
 \begin{pmatrix}
     c_n   		& a_n	        	& 0		  \\
     a_n^*			& 0			& b_n		  \\
     0   		& b_n^*	 		& 0\\
 \end{pmatrix},
 \hspace{0.5cm}
 M'_\ell=
   \begin{pmatrix}
    0   		& a_{\ell} 	& 0			\\
    a_{\ell}^*		& d_{\ell}			& b_{\ell}	\\
    0		   	& b_{\ell}^* 	 	& c_{\ell}			\\
    \end{pmatrix}.
\end{align}
In what follows we are going to analyze the consequences of this particular pattern 
with three texture zeros in the neutral sector and two in the charged one.

The first step is to remove the phases; this can be done by the following unitary 
transformation: 

\begin{align}
M'_{n,\ell}
=\lambda_{n,\ell}^\dagger
M_{n,\ell}
  \lambda_{n,\ell},
% \notag\\ 
 \end{align}
\noindent
which is achieved by using the diagonal matrices 
$\lambda_n=(1,e^{i\alpha_{n_1}},e^{i\alpha_{n_1}+i\alpha_{n_2}})$
and $\lambda_\ell =(1,e^{i\alpha_{\ell_1}},e^{i\alpha_{\ell_1}+i\alpha_{\ell_2}})$, 
respectively, and $ M_ {n,\ell} $ are the matrices whose components are the absolute values of the corresponding entries in $ M_ {n,\ell}^{\prime}$
(i.e., $(M_ {n,\ell})_{i,j} = |(M_ {n,\ell}^{\prime})_{i,j}|$).
If we rotate these matrices by using the orthogonal transformation 
$R_{n,\ell}\:(R_{n,\ell}^T\, R_{n,\ell}= \mathbbm{1})$ to the mass eigenstate 
space (the physical basis), we get
\begin{align}
M'_{n,\ell}
= \lambda_{n,\ell}^\dagger R^{T}_{n,\ell} 
  \begin{pmatrix}
   m_{1,e}  &  0   & 0  \\
     0  & -m_{2,\mu} & 0  \\
     0  &  0   & m_{3,\tau}
 \end{pmatrix} 
 R_{n,\ell}
 \lambda_{n,\ell}
 \equiv U_{n,\ell} M^{\text{diag}}_{n,\ell} U^{\dagger}_{n,\ell},
 \end{align}
where at least  a negative eigenvalue is needed in order to generate a texture-zero in 
the diagonal~\cite{Branco:1999nb,Fusaoka:1998vc,Giraldo:2011ya,Giraldo:2015cpp}. Here 
$m_1,m_2$ and $m_3$ are the masses of the electron, muon and tau neutrinos 
respectively, with the masses of the charged leptons given by:
%\begin{align}
  $m_e=0.5109989461\pm0.0000000031,
m_\mu= 105.6583745\pm0.0000024$ 
and $ m_\tau= 1776.86\pm0.12,$
%\end{align}
which correspond to the electron,  muon and 
tau masses, respectively~(in MeV)~\cite{Tanabashi:2018oca}.
After rotating to the mass eigenstates the eigenvalues can be positive, negative or zero.
In these expressions $M_n^{\text{diag}}$ and $M_\ell^{\text{diag}}$ are  
the diagonal mass matrices for the neutrino and charged lepton  
sectors respectively.  In accord with the standard notation we use  
$U_n\equiv (R_n\lambda_n)^{\dagger}$ and $U_\ell\equiv (R_\ell\lambda_\ell)^{\dagger}$, 
two unitary matrices used to rotate from the weak basis to the physical basis.
From equations~\eqref{eq:dmass} and \eqref{eq:un} we obtain the relation 
between the states in the mass basis $\nu_{L,R}$, $\ell_{L,R}$ and the corresponding 
states in the interaction basis $\nu'_{L,R}$, $\ell'_{L,R}$:
\begin{align}
\nu'_{L,R}= U_n \nu_{L,R},\hspace{1cm}
 \ell'_{L,R}=  U_\ell \ell_{L,R}.
\end{align}
Replacing these expressions in the lepton sector of the weak current we obtain
\begin{align}
%J^{W\mu}_L
\mathcal{L}_{W^{-}}= -\frac{g}{\sqrt{2}}W^{-}\bar{\ell}'_L \gamma^{\mu}\nu'_{L}+\text{h.c}
                     = -\frac{g}{\sqrt{2}}W^{-}\bar{\ell}_L \gamma^{\mu}U_\ell^{\dagger}U_n\nu_{L}+\text{h.c},
\end{align}
in such a way that the Pontecorvo-Maki-Nakagawa-Sakata matrix (PMNS matrix) is given by 
\begin{align}\label{eq:pmns}
V_{\text{PMNS}}=U_\ell^{\dagger}U_n % R^{I,II}_{\ell}\lambda_\ell\lambda_n^{\dagger} (R_n^{I,II})^T
                                   = R_{\ell}\:\Phi\:R_n^T,
     \end{align}
where $\Phi=\lambda_\ell\lambda_n^{\dagger}$ is a diagonal phase matrix.
For the neutrino mass matrix normal ordering is assumed~\cite{Verma:2016qhy}, i.e.:
$m_3>m_2>m_1$,  where:
$m^2_2 = m^2_1+\delta m^2_{21}$, 
and 
$m^2_3 = m^2_1+\delta m^2_{31}$,  with $\delta m^2_{21},
\delta m^2_{31}>0$~\cite{Esteban:2018azc}.

By imposing the invariance of the trace and the determinant of the mass matrices 
($\text{tr}[M'_{n,\ell}]=\text{tr}[M_{n,\ell}^{\text{diag}}]$, 
$\text{tr}\left[\left(M_{n,\ell}^{\prime}\right)^2\right]=
\text{tr}\left[\left(M_{n,\ell}^{\text{diag}}\right)^2\right]$, 
and  $\text{Det}[M_{n,\ell}^\prime]=\text{Det}[M_{n,\ell}^{\text{diag}}]$), 
the following relations are obtained for this particular texture:

\begin{align}
\begin{split}
c_{n}&=m_1-m_2+m_3,\\
|a_{n}|&=\sqrt{\frac{(m_1-m_2)(m_1+m_3)(m_2-m_3)}{m_1-m_2+m_3}},\\
|b_{n}|&=\sqrt{\frac{m_1\,m_2\,m_3}{m_1-m_2+m_3}},
\end{split}
\begin{split}
d_{\ell}&=m_e-m_\mu+m_\tau-c_\ell,\nonumber\\
|b_{\ell}|&=\sqrt{\frac{(c_\ell-m_e)(c_\ell+m_\mu)(m_\tau-c_\ell)}{c_\ell}},\nonumber\\
|a_{\ell}|&=\sqrt{\frac{m_e\,m_\mu\,m_\tau}{c_\ell}}.\nonumber
\end{split}
\end{align}
From the previous identifications, it is possible to obtain an explicit form for 
the mass matrices of leptons that allows us to obtain, through diagonalization of 
$M_n$ and $M_{\ell}$, the orthogonal matrices in Eq.~\eqref{eq:pmns}, 

\begin{align}
%	{\cal O}_{n}=
\label{eqA1}
R_n&= 
	\begin{pmatrix}
   - \sqrt{\frac{m_1\left(m_2-m_1\right)\left(m_1+m_3\right)}{\left(m_1+m_2\right)\left(m_3-m_1\right)\left(m_1-m_2+m_3\right)}}	&  	\sqrt{\frac{m_1\left(m_3-m_2\right)}{\left(m_1+m_2\right)\left(m_3-m_1\right)}}		& 	\sqrt{\frac{m_2 m_3\left(m_3-m_2\right)}{\left(m_1+m_2\right)\left(m_3-m_1\right)\left(m_1-m_2+m_3\right)}}	  \\
    \sqrt{\frac{m_2\left(m_1-m_2\right)\left(m_2-m_3\right)}{\left(m_1+m_2\right)\left(m_2+m_3\right)\left(m_1-m_2+m_3\right)}}	&  	-\sqrt{\frac{m_2\left(m_1+m_3\right)}{\left(m_1+m_2\right)\left(m_2+m_3\right)}}		& 	\sqrt{\frac{m_1 m_3\left(m_1+m_3\right)}{\left(m_1+m_2\right)\left(m_2+m_3\right)\left(m_1-m_2+m_3\right)}}	  \\
    \sqrt{\frac{m_3\left(m_1+m_3\right)\left(m_3-m_2\right)}{\left(m_3-m_1\right)\left(m_2+m_3\right)\left(m_1-m_2+m_3\right)}}	&      \sqrt{\frac{m_3\left(m_2-m_1\right)}{\left(m_2+m_3\right)\left(m_3-m_1\right)}}		& 	\sqrt{\frac{m_1 m_2\left(m_2-m_1\right)}{\left(m_3-m_1\right)\left(m_2+m_3\right)\left(m_1-m_2+m_3\right)}}	  \\
	\end{pmatrix},\\
\label{eqA2}
R_\ell&= 
	\begin{pmatrix}
  -\sqrt{\frac{m_\mu  m_\tau  (c_\ell-m_e)}{c_\ell (me+m_\mu )
   (m_\tau -m_e)}}	&  	-\sqrt{\frac{m_e (c_\ell-m_e)}{(m_e+m_\mu ) (m_\tau -m_e)}}	& \sqrt{\frac{m_e (c_\ell+m_\mu ) (c_\ell-m_\tau)}{c_\ell (m_e+m_\mu )
   (m_e-m_\tau) }}	  \\
   \sqrt{\frac{m_em_\tau(c_\ell+m_\mu)}{c_\ell(m_e+m_\mu)(m_\mu+m_\tau)}}	&  	-\sqrt{\frac{m_\mu  (c_\ell+m_\mu )}{(m_e+m_\mu ) (m_\mu +m_\tau
   )}}		& 	\sqrt{\frac{m_\mu  (m_e-c_\ell) (c_\ell-m_\tau )}{c_\ell (m_e+m_\mu )
   (m_\mu +m_\tau )}}	  \\
  \sqrt{\frac{m_e m_\mu  (c_\ell-m_\tau )}{c_\ell (m_e-m_\tau )
   (m_\mu +m_\tau )}}	& \sqrt{\frac{m_\tau  (m_\tau -c_\ell)}{(m_\tau -m_e) (m_\mu +m_\tau
   )}}		& 	\sqrt{\frac{m_\tau  (c_\ell-m_e) (c_\ell+m_\mu )}{c_\ell (m_\tau -m_e)
   (m_\mu +m_\tau )}}	  \\
	\end{pmatrix}.
\end{align}
The  entries of ${R}_{n}$ and in  ${R}_{\ell}$ are real values because of the normal hierarchy 
assumed in neutrino masses, and to the hierarchy of the charged lepton 
sector $m_\tau > m_\mu > m_e$, with this in mind the  
constraint for $c_\ell$ is established to be in the interval $ m_e<c_\ell<m_\tau $. 
A closer look to the former expressions shows that for this analysis $m_1$ and $c_\ell$ 
can be taken as free parameters to be fixed by an statistical analysis.
%%%%%%%%%%%%%%%%%%%%%%%%%%%%%%%%%%%%%%%%%%%%%%%%%%%%%%%%%%%%
%%%%%%%%%%%%%%%%%%%%%%%%%%%%%%%%%%%%%%%%%%%%%%%%%%%%%%%%%%%%
\subsection{Least squares analysis}
From equation \eqref{eq:pmns} we have that 
$V_{\rm PMNS}=  R_{\ell}\,\Phi\,R_{n}^{T}$, with $\Phi$ the following diagonal matrix:
\begin{equation}\nonumber
\Phi=
\begin{pmatrix}
    1		  	& 0		 		& 0			 	\\
    0			& e^{i \phi_1}	& 0				\\
    0		   	& 0		 	 	& e^{i \phi_2}	\\
\end{pmatrix},
\end{equation}
where $\phi_1$ and $\phi_2$ are free parameters. 

Then, the former analysis imply that the matrix $V_{\text{PMNS}}$ is a function 
of the free parameters  $(m_1,c_\ell,\phi_1,\phi_2)$ 
where we have chosen  $m_1$ as the lightest neutrino mass.
After a numerical adjustment by means of the $\chi^2$ analysis, it is found that 
for our particular choice of the five-zero texture lepton mass matrices in  
Eq.~\eqref{eq:un}, the normal hierarchy is favored. 
Neglecting correlations, the $\chi^{2}$ function is given by 
\begin{equation}\nonumber
\chi^{2}=P^{2}_{J}+\sum\limits_{i,j=1,2,3}P^{2}_{ij},	
\end{equation}
where the pulls are
\begin{equation}\nonumber
P_{ij}=\frac{U_{ij}-\bar{U}_{ij}}{\delta U_{ij}} 
\end{equation}
where  $U_{ij}=|(V_{\rm PMNS})_{ij}|$ is the absolute value of the components from the product of the  diagonalization matrices~(\ref{eq:pmns}), the absolute values  $\bar{U}_{ij}$ correspond to the global averages for the components of the PMNS matrix and $\delta U_{ij}$  corresponding 1$\sigma$ errors. 
$P_J$ is the  pull of the Jarlskog invariant  
which, in the standard parameterization, is given by
 $\bar{J}=c_{12}c_{23}c^{2}_{13}s_{12}s_{23}s_{13}\sin{\delta}=-0.0270054$ and the corresponding 1$\sigma$ uncertainty is
 $\delta J=0.0106304$, for normal ordering~\cite{Tanabashi:2018oca}. The theoretical prediction is given by $J={\rm Im} \left(U_{\mu 3}U^{*}_{\tau 3}U_{\mu 2}U^{*}_{\tau 2}\right)$,
 where in this expression $U$ stands for the PMNS mixing matrix.   The upper bound 
$m_1+m_2+m_3<0.17$\,\,\, eV~\cite{Tanabashi:2018oca}, is also imposed. Using the data from~\cite{Esteban:2018azc}~\footnote{NuFIT collaboration~(http://www.nu-fit.org/?q=node/211)(with SK atmospheric data).},
the fit results are shown in the following tables:
\begin{table}[ht]
\begin{center}
\begin{turn}{0}
\begin{tabular}{| c | c | c | c | c | c | c | c | c | }
\hline
 $m_1\left({\rm eV}\right)$  & $c_{\ell}\left({\rm eV}\right)$ & $\phi_{1} \left({\rm rad}\right)$  & $\phi_{2}\left({\rm rad}\right)$ & $\chi^{2}_{\min}$  \\ \hline
%\hline \hline
%
$0.00395\pm^{0.00062}_{0.00078}$ & 523176. & 0.0190664 & 1.56122 & 12.4204 \\ \hline 
\end{tabular}
\end{turn}
\caption{Best fit free parameters and minimum $\chi^2$ function.} 
\label{table4}
\end{center}
\end{table}

%%%%%%%%%%%%%%%%%%%%%%%%%%%%%%%%%%%%%%%%%%%%%%%%%%%%

%\begin{spacing}{}

\begin{table}[ht]
\begin{center}
\begin{turn}{0}
\begin{tabular}{| c | c | c | c | c | c | c | c | c | c |}
\hline
 $P_{11}$  &  $P_{12}$  & $P_{13}$ & $P_{21}$  & $P_{22}$ & $P_{23}$ & $P_{31}$ & $P_{32}$ & $P_{33}$ & $P_{J}$\\ \hline
0.428531 & -0.385085 & 0.0430767 & -0.205321 & -1.2577 & 1.91336 & 0.290472 & 1.25036  & -2.2701 & 0.0228083 \\ \hline 
\end{tabular}
\end{turn}
\caption{$P_{i,j}$  is the pull of the $i, j$ component of the PMNS matrix and $P_J$ is the pull of  the Jarlskog invariant in the $\chi^2$ analysis.  The minimum of the $\chi^2$ function is $12.4204$ for ten observables and four parameters $\left(m_1, c_{\ell},\phi_{1},\phi_{2}\right)$. The fit goodness is $\chi^2/\text{d.o.f}=2.07$ which is a relatively high value due to $P_{33}$ and  $P_{23}$  pulls which have a deviation around  $2\sigma$ respect to their experimental values, despite this result it is still an acceptable  fit. }
\label{table5}
\end{center}
\end{table}
%\end{spacing}
In our $\chi^2$  analysis the pseudo observables are the absolute values of the PMNS matrix components and the Jarlskog invariant, with pulls  $P_{i,j}$ and $P_J$, respectively. We do not consider correlations between them~\footnote{The collaborations report correlation effects between observables, in our case, the components of the PMNS matrix are the result of a global fit. However, in phenomenology is a common practice to use pseudo observables.}. \\

Even though that a value for the fit goodness  $\chi^2/\text{d.o.f}\sim 2.07$ is not optimal, the result is acceptable.  We can see that the main source of tension is related to the $(V_{\text{PMNS}})_{23}$  and  $(V_{\text{PMNS}})_{33}$ components, which deviate from their experimental values by 2$\sigma$.  It is important to stress that a lightest neutrino mass equal to zero  and the inverse ordering of the neutrino masses,   they  are not favored by this texture~(the same is true for the equivalent textures via weak basis transformations). It is possible that for another non-equivalent five-zero texture a realization of the inverse ordering.
This subject requires a more dedicated study.

%%%%%%%%%%%%%%%%%%%%%%%%%%%%%%%%%%%%%%%%%%%%%%%%%%%%%%%%%%%%%%%%%%%%%%%%%%
%%%%%%%%%%%%%%%%%%%%%%%%%%%%%%%%%%%%%%%%%%%%%%%%%%%%%%%%%%%%%%%%%%%%%%%%%%
\FloatBarrier
\section{Five texture zeros. Second case.}
Let us now assume that in the context of the SMRHN with the neutrinos being only 
Dirac type particles, there is a symmetry which produces the  
Hermitian mass matrices $M'_n$ and $M_\ell'$ with the following textures
\begin{align}\label{eq:unp}
 M'_n=
 \begin{pmatrix}
     0  		& C_n	        	& 0		  \\
     C_n^*		& D_n			& B_n		  \\
     0   		& B_n^*	 		& A_n\\
 \end{pmatrix},
 \hspace{0.5cm}
 M'_\ell=
   \begin{pmatrix}
    0   		& C_{\ell} 	     & 0			\\
    C_{\ell}^*		& 0		     & B_{\ell}	\\
    0		   	& B_{\ell}^* 	     & A_{\ell}			\\
    \end{pmatrix}.
\end{align}
Let us analyze the consequences of this new pattern 
with three texture zeros in the charged sector and two in the neutral one. Without losing generality, it is possible to eliminate the phases of the $M'_{\ell} $ matrix by means of a WBT, so that the CP violation phase only appears in the neutrino mass matrix.
The algebra shows that it is possible to diagonalize the lepton sector 
$M_\ell=U_\ell D_\ell U_\ell^\dagger$~(which as in the previous section we define 
$(M_{\ell})_{i,j}= |(M'_{\ell})_{i,j}|$), where $D_{\ell}=$Diag.$(m_e,-m_\mu,m_\tau)$, in order to take full advantage of using the 
following unitary matrix
\begin{widetext}
\begin{equation} 
\label{32}
{%\footnotesize
 U_\ell=\begin{pmatrix}
     e^{i\theta_1}\, 
\sqrt{\frac{m_\mu m_\tau(A_\ell-m_e)}{A_\ell(m_\mu
+m_e)(m_\tau-m_e)}}&-e^{i\theta_2}
\sqrt{\frac{m_em_\tau(m_\mu+A_l)}{A_\ell(m_\mu
+m_e)(m_\tau+m_\mu)}}&
\sqrt{\frac{-m_em_\mu(A_\ell-m_\tau)}{A_\ell(m_\tau
-m_e)(m_\tau+m_\mu)}}\\
&&&\\[-2mm]
e^{i\theta_1}\sqrt{\frac{m_e(m_e-A_\ell)}{(-m_\mu
-m_e)(m_\tau-m_e)}}&
e^{i\theta_2}\sqrt{\frac{m_\mu(A_\ell+m_\mu)}{(m_\mu+m_e )(m_\tau+m_\mu)}}&
\sqrt{\frac{m_\tau(m_\tau-A_\ell)}{(m_\tau-m_e
)(m_\tau+m_\mu)}}\\
&&&\\[-2mm]
-e^{i\theta_1}\sqrt{\frac{m_e(A_l+m_\mu)(A_\ell-m_\tau)}
{ A_\ell(-m_\mu-m_e)(m_\tau-m_e)}}&
-e^{i\theta_2}\sqrt{\frac{m_\mu(A_\ell-m_e)(m_\tau-A_\ell)}{
A_\ell(m_\mu+m_e)(m_\tau+m_\mu)}}&
\sqrt{\frac{m_\tau(A_l-m_e)(A_\ell+m_\mu)}{A_\ell(m_\tau
-m_e)(m_\tau+m_\mu)}}
     \end{pmatrix},}%\quad \text{or}\\ % \nonumber 
\end{equation}
\end{widetext}
where $\theta_1$ and  $\theta_2$ are arbitrary phases and 
  $A_\ell=m_e-m_\mu+m_\tau$.
Even though the phases, $\theta_1$ and $\theta_2$, in the rotation matrix of $U_l$ are no CP phases (they can be absorbed in the fields),  these phases are quite useful to match our theoretical expression for the PMNS matrix  with the standard convention~\cite{Rasin:1997pn}.
To get the three texture zeros in the lepton mass matrix the following  relations are also necessary
\begin{align*}
|B_\ell|&=\sqrt{\frac{(A_\ell-m_e)(A_\ell+m_\mu)(m_\tau
-A_\ell)}{A_\ell}}\qquad\text{and}\qquad
|C_\ell|=\sqrt{\frac{m_e\,m_\mu\,m_\tau}{A_\ell}}.
\end{align*}

For the neutrino sector we are subject to the condition
$U_\ell^{\dagger}U_n= V_{\text{\tiny PMNS}}$ and, necessarily, the diagonalizing matrix 
must be given by $U_n=U_{\ell}V_{\text{\tiny PMNS}}$,
so that the relation between the mass matrix in the weak basis and the 
diagonal matrix $D_n$ in the mass space is
%\begin{widetext}
\begin{align}
\label{eq15}
 M_n^\prime&=
\begin{pmatrix}
  0 &C_n& 0\\
C_n^*& D_n& B_n\\
0& B_n^* & A_n
\end{pmatrix}=
U_\ell(V_{\text{\tiny 
PMNS}})D_n(V_{\text{\tiny PMNS}})^\dag U_\ell^\dag\equiv U_n D_n U_n^{\dagger}.
     \end{align}
%
%\end{widetext}
%
For this second case the only free parameters are: $m_1$ from the diagonal matrix 
$D_n$=Diag.$(m_1,-m_2,m_3)$, and $\theta_1$ and $\theta_2$ from $U_{\ell}$. 
This result is  important since we can interpret the neutrino masses as predictions 
associated with the texture of the mass matrices. 
From these expressions we can obtain useful relations 
%as those in the equation~(\ref {1.1}) 
by identifying  $U_{\ell}$  with  the weak basis transformation $U$ in 
equation~(\ref{wbss})~\cite{Giraldo:2011ya}.

\subsection{Numerical results}
For this second case, when solving numerically to obtain the texture for the neutrino 
mass matrix in the normal hierarchy for the neutrino  masses, we obtain
\begin{equation}
 \begin{split}
  m_1&=(0.00354\pm0.00088)\,\text{eV},\\
  m_2&=(0.00930\pm0.00036)\,\text{eV},\\
  m_3&=(0.05040\pm0.00030) \,\text{eV}.
 \end{split}
\end{equation}
In our numerical analysis the main source of uncertainty comes from the CP 
violation phase, result which makes sense since in the lepton sector this 
parameter has not been determined with good precision.
The associated numerical entries for lepton mass matrices with five texture 
zeros  are (in MeV): $A_n= 0.0251821$, $B_n= (-0.0122955+0.0244187 i)$, 
$C_n=(0.00427236\, +0.00689527 i)$, $D_n=0.0194623$, 
$A_\ell=1671.71$, $|B_\ell|=432.237$, $|C_\ell|=7.57544$, and phases  $\theta_1=0.154895$ and $\theta_2= 2.01797$. The phases of $B_{\ell}$ and $C_{\ell}$ were absorbed in $B_{n}$ and $C_{n}$ by means of a redefinition, through a WBT,  in a previous step.
\begin{widetext}
By construction, the WBT formalism reproduces the mixing matrix,  the mass of 
charged leptons and the neutrino mass squared differences. We used, as input parameters, the central values of the global fit reported by the Nu-FIT collaboration~(with SK atmospheric data)~\cite{Esteban:2018azc}.
When comparing with the method of least squares in the WBT there are no 
deviations from the experimental values, see Table~\ref{tab:input}.
\bigskip
\begin{center}
\begin{table}[h]
 \begin{tabular}{|c|c|c|c|c|c|c|c|c|}
 \hline
  $\theta_{12}\,(^\circ)$& $\theta_{23}\,(^\circ)$& 
$\theta_{13}\,(^\circ)$&$\delta_{CP}\,(^\circ)$&$\delta 
m_{21}^2\,(eV^2)$&$\delta 
m_{31}^2\,(eV^2)$&$m_e\,(\text{MeV})$&$m_\mu\,(\text{MeV})$&$m_\tau\,(\text{MeV}
)$\\[1mm]\hline
&&&&&&&&\\
33.82&48.6& 8.60&221 & 
$7.39\times10^{-5}$&$2.528\times10^{-3}$ &0.510999 &105.658
&1776.86\\
\hline 
\end{tabular}
\label{tab:input}
\caption{Output values in our analysis. }
\end{table}
\end{center}

\end{widetext}
\section{Conclusions}
In this analysis, we explore the consequences of extending the SM with three right-handed neutrinos that allow nine additional complex Dirac mass terms for the neutral lepton sector, and excluding the possibility of having bare Majorana masses. 
Two different five texture zeros for the Hermitian lepton mass matrices were 
considered, one with three zeros in the neutral sector and two in the charged sector, 
and the other one with two texture zeros in the neutral sector and three in the 
charged one. In order to have reliable results we have used two different approaches 
assuming for both a normal ordering for the neutrino physical masses. 

Counting the degrees of freedom in the lepton sector, and after making use of  
the polar theorem of matrix algebra and the consequences of the WBT, 
we have concluded that with five texture zeros in the Hermitian mass matrices, 
only one prediction can be reached. We have chosen the lightest neutrino mass 
for such a predicted value.

The first analysis, based on a least squares approach used to adjust the lepton 
masses and the mixing parameters to their 
corresponding experimental values, was implemented for the texture with three zeros 
in the neutral sector and two in the charged one.
In this approach, the fit of the mixing parameters 
to the reported values in the literature is below two sigmas with an acceptable 
fit goodness. The best fit for the lightest neutrino mass in this case was  
$m_{1}\approx (3.9\pm^{0.6}_{0.8})\times 10^{-3}$\,eV, which is 
similar to reported values starting from other 
assumptions~\cite{Fritzsch:2015gxa,Fritzsch:2015haa,Fritzsch:2016xmb}.

The second analysis was a purely algebraic and numerical study, based on the WBT approach. It was implemented 
for the texture with three zeros in the charged lepton sector and two in the 
neutral one. The prediction now for the lightest neutrino mass was 
$(3.5\pm0.9)\times 10^{-3}$~eV., in agreement with the previous result.

The two different five texture-zeros proposed in expressions~\eqref{eq:un} 
and~\eqref{eq:unp} are not equivalent in the sense that there is not a WBT 
relating them.

\section*{Acknowledgments}
This work was partially supported by VIPRI in the Universidad de Nariño, Approval Contract No. 024 and 160, Project Code: 1048 and 1928. R.H.B and L.M. thanks the ''Centro de Laboratorios de investigación parque i-ITM''.

\FloatBarrier

%\appendix

%\bibliographystyle{apsrev4-1}
%\bibliographystyle{apsrev4-1longdoi}
%\bibliographystyle{apsrev4-1long}
%\bibliography{references}
%\bibliography{prueba}

\begin{thebibliography}{99}
%\cite{Donoghue:1992dd}
\bibitem{Donoghue:1992dd} 
  J.~F.~Donoghue, E.~Golowich and B.~R.~Holstein,
  %``Dynamics of the standard model,''
  Camb.\ Monogr.\ Part.\ Phys.\ Nucl.\ Phys.\ Cosmol.\  {\bf 2}, 1 (1992)
  [Camb.\ Monogr.\ Part.\ Phys.\ Nucl.\ Phys.\ Cosmol.\  {\bf 35} (2014)].
  doi:10.1017/CBO9780511524370
  %%CITATION = doi:10.1017/CBO9780511524370;%%
  %303 citations counted in INSPIRE as of 05 Feb 2020



%\cite{Peinado:2019mrn}
\bibitem{Peinado:2019mrn} 
  E.~Peinado, M.~Reig, R.~Srivastava and J.~W.~F.~Valle,
  %``Dirac neutrinos from Peccei-Quinn symmetry: a fresh look at the axion,''
  arXiv:1910.02961 [hep-ph].
  %%CITATION = ARXIV:1910.02961;%%
  %4 citations counted in INSPIRE as of 05 Feb 2020



%\cite{Minkowski:1977sc}
\bibitem{Minkowski:1977sc} 
  P.~Minkowski,
  %``$\mu \to e\gamma$ at a Rate of One Out of $10^{9}$ Muon Decays?,''
  Phys.\ Lett.\  {\bf 67B}, 421 (1977).
  doi:10.1016/0370-2693(77)90435-X
  %%CITATION = doi:10.1016/0370-2693(77)90435-X;%%
  %3662 citations counted in INSPIRE as of 05 Feb 2020



%\cite{GellMann:1980vs}
\bibitem{GellMann:1980vs} 
  M.~Gell-Mann, P.~Ramond and R.~Slansky,
  %``Complex Spinors and Unified Theories,''
  Conf.\ Proc.\ C {\bf 790927}, 315 (1979)
  [arXiv:1306.4669 [hep-th]].
  %%CITATION = ARXIV:1306.4669;%%
  %3045 citations counted in INSPIRE as of 05 Feb 2020



%\cite{Yanagida:1980xy}
\bibitem{Yanagida:1980xy} 
  T.~Yanagida,
  %``Horizontal Symmetry and Masses of Neutrinos,''
  Prog.\ Theor.\ Phys.\  {\bf 64}, 1103 (1980).
  doi:10.1143/PTP.64.1103
  %%CITATION = doi:10.1143/PTP.64.1103;%%
  %575 citations counted in INSPIRE as of 05 Feb 2020



%\cite{Glashow:1979nm}
\bibitem{Glashow:1979nm} 
  S.~L.~Glashow,
  %``The Future of Elementary Particle Physics,''
  NATO Sci.\ Ser.\ B {\bf 61}, 687 (1980).
  doi:10.1007/978-1-4684-7197-7\_15
  %%CITATION = doi:10.1007/978-1-4684-7197-7_15;%%
  %392 citations counted in INSPIRE as of 05 Feb 2020



%\cite{Mohapatra:1979ia}
\bibitem{Mohapatra:1979ia} 
  R.~N.~Mohapatra and G.~Senjanovic,
  %``Neutrino Mass and Spontaneous Parity Nonconservation,''
  Phys.\ Rev.\ Lett.\  {\bf 44}, 912 (1980).
  doi:10.1103/PhysRevLett.44.912
  %%CITATION = doi:10.1103/PhysRevLett.44.912;%%
  %5223 citations counted in INSPIRE as of 05 Feb 2020



%\cite{Schechter:1980gr}
\bibitem{Schechter:1980gr} 
  J.~Schechter and J.~W.~F.~Valle,
  %``Neutrino Masses in SU(2) x U(1) Theories,''
  Phys.\ Rev.\ D {\bf 22}, 2227 (1980).
  doi:10.1103/PhysRevD.22.2227
  %%CITATION = doi:10.1103/PhysRevD.22.2227;%%
  %2708 citations counted in INSPIRE as of 05 Feb 2020



%\cite{Ahuja:2009jj}
\bibitem{Ahuja:2009jj} 
  G.~Ahuja, M.~Gupta, M.~Randhawa and R.~Verma,
  %``Texture specific mass matrices with Dirac neutrinos and their implications,''
  Phys.\ Rev.\ D {\bf 79}, 093006 (2009)
  doi:10.1103/PhysRevD.79.093006
  [arXiv:0904.4534 [hep-ph]].
  %%CITATION = doi:10.1103/PhysRevD.79.093006;%%
  %31 citations counted in INSPIRE as of 05 Feb 2020



%\cite{Liu:2012axa}
\bibitem{Liu:2012axa} 
  X.~w.~Liu and S.~Zhou,
  %``Texture Zeros for Dirac Neutrinos and Current Experimental Tests,''
  Int.\ J.\ Mod.\ Phys.\ A {\bf 28}, 1350040 (2013)
  doi:10.1142/S0217751X13500401
  [arXiv:1211.0472 [hep-ph]].
  %%CITATION = doi:10.1142/S0217751X13500401;%%
  %16 citations counted in INSPIRE as of 05 Feb 2020



%\cite{Verma:2013cza}
\bibitem{Verma:2013cza} 
  R.~Verma,
  %``Lower bound on neutrino mass and possible CP violation in neutrino oscillations,''
  Phys.\ Rev.\ D {\bf 88}, 111301 (2013)
  doi:10.1103/PhysRevD.88.111301
  [arXiv:1311.7524 [hep-ph]].
  %%CITATION = doi:10.1103/PhysRevD.88.111301;%%
  %6 citations counted in INSPIRE as of 05 Feb 2020



%\cite{Ludl:2014axa}
\bibitem{Ludl:2014axa} 
  P.~O.~Ludl and W.~Grimus,
  %``A complete survey of texture zeros in the lepton mass matrices,''
  JHEP {\bf 1407}, 090 (2014)
  Erratum: [JHEP {\bf 1410}, 126 (2014)]
  doi:10.1007/JHEP07(2014)090, 10.1007/JHEP10(2014)126
  [arXiv:1406.3546 [hep-ph]].
  %%CITATION = doi:10.1007/JHEP07(2014)090, 10.1007/JHEP10(2014)126;%%
  %45 citations counted in INSPIRE as of 05 Feb 2020



%\cite{Verma:2014lpa}
\bibitem{Verma:2014lpa} 
  R.~Verma,
  %``Lepton textures and neutrino oscillations,''
  Int.\ J.\ Mod.\ Phys.\ A {\bf 29}, no. 21, 1444009 (2014)
  doi:10.1142/S0217751X14440096
  [arXiv:1406.0640 [hep-ph]].
  %%CITATION = doi:10.1142/S0217751X14440096;%%
  %4 citations counted in INSPIRE as of 05 Feb 2020



%\cite{Fakay:2014nea}
\bibitem{Fakay:2014nea} 
  P.~Fakay, S.~Sharma, G.~Ahuja and M.~Gupta,
  %``Leptonic mixing angle $\theta_{13}$ and ruling out of minimal texture for Dirac neutrinos,''
  PTEP {\bf 2014}, no. 2, 023B03 (2014)
  doi:10.1093/ptep/ptu005
  [arXiv:1401.8121 [hep-ph]].
  %%CITATION = doi:10.1093/ptep/ptu005;%%
  %3 citations counted in INSPIRE as of 05 Feb 2020



%\cite{Cebola:2015dwa}
\bibitem{Cebola:2015dwa} 
  L.~M.~Cebola, D.~Emmanuel-Costa and R.~G.~Felipe,
  %``Confronting predictive texture zeros in lepton mass matrices with current data,''
  Phys.\ Rev.\ D {\bf 92}, no. 2, 025005 (2015)
  doi:10.1103/PhysRevD.92.025005
  [arXiv:1504.06594 [hep-ph]].
  %%CITATION = doi:10.1103/PhysRevD.92.025005;%%
  %25 citations counted in INSPIRE as of 05 Feb 2020



%\cite{Gautam:2015kya}
\bibitem{Gautam:2015kya} 
  R.~R.~Gautam, M.~Singh and M.~Gupta,
  %``Neutrino mass matrices with one texture zero and a vanishing neutrino mass,''
  Phys.\ Rev.\ D {\bf 92}, no. 1, 013006 (2015)
  doi:10.1103/PhysRevD.92.013006
  [arXiv:1506.04868 [hep-ph]].
  %%CITATION = doi:10.1103/PhysRevD.92.013006;%%
  %29 citations counted in INSPIRE as of 05 Feb 2020



%\cite{Ahuja:2016san}
\bibitem{Ahuja:2016san} 
  G.~Ahuja, S.~Sharma, P.~Fakay and M.~Gupta,
  %``General lepton textures and their implications,''
  Mod.\ Phys.\ Lett.\ A {\bf 30}, no. 34, 1530025 (2015)
  doi:10.1142/S0217732315300256
  [arXiv:1604.03339 [hep-ph]].
  %%CITATION = doi:10.1142/S0217732315300256;%%
  %5 citations counted in INSPIRE as of 05 Feb 2020



%\cite{Singh:2018lao}
\bibitem{Singh:2018lao} 
  M.~Singh,
  %``Texture One Zero Dirac Neutrino Mass Matrix With Vanishing Determinant or Trace Condition,''
  Nucl.\ Phys.\ B {\bf 931}, 446 (2018)
  doi:10.1016/j.nuclphysb.2018.05.004
  [arXiv:1804.00842 [hep-ph]].
  %%CITATION = doi:10.1016/j.nuclphysb.2018.05.004;%%
  %1 citations counted in INSPIRE as of 05 Feb 2020



%\cite{Ahuja:2018fmw}
\bibitem{Ahuja:2018fmw} 
  G.~Ahuja and M.~Gupta,
  %``Texture zero mass matrices and nature of neutrinos,''
  Int.\ J.\ Mod.\ Phys.\ A {\bf 33}, no. 31, 1844032 (2018).
  doi:10.1142/S0217751X18440323
  %%CITATION = doi:10.1142/S0217751X18440323;%%



%\cite{DellOro:2016tmg}
\bibitem{DellOro:2016tmg} 
  S.~Dell'Oro, S.~Marcocci, M.~Viel and F.~Vissani,
  %``Neutrinoless double beta decay: 2015 review,''
  Adv.\ High Energy Phys.\  {\bf 2016}, 2162659 (2016)
  doi:10.1155/2016/2162659
  [arXiv:1601.07512 [hep-ph]].
  %%CITATION = doi:10.1155/2016/2162659;%%
  %252 citations counted in INSPIRE as of 05 Feb 2020



%\cite{Fritzsch:2015gxa}
\bibitem{Fritzsch:2015gxa} 
  H.~Fritzsch,
  %``Neutrino Masses and Flavor Mixing,''
  Mod.\ Phys.\ Lett.\ A {\bf 30}, no. 16, 1530012 (2015)
  doi:10.1142/S0217732315300128
  [arXiv:1503.01857 [hep-ph]].
  %%CITATION = doi:10.1142/S0217732315300128;%%
  %8 citations counted in INSPIRE as of 05 Feb 2020



%\cite{Fritzsch:2015haa}
\bibitem{Fritzsch:2015haa} 
  H.~Fritzsch,
  %``Texture Zero Mass Matrices and Flavor Mixing of Quarks and Leptons,''
  Subnucl.\ Ser.\  {\bf 53}, 201 (2017)
  doi:10.1142/9789813208292\_0006
  [arXiv:1503.07927 [hep-ph]].
  %%CITATION = doi:10.1142/9789813208292_0006;%%
  %4 citations counted in INSPIRE as of 05 Feb 2020



%\cite{Fritzsch:2016xmb}
\bibitem{Fritzsch:2016xmb} 
  H.~Fritzsch,
  %``Neutrino oscillations and neutrino masses,''
  Mod.\ Phys.\ Lett.\ A {\bf 31}, no. 15, 1630014 (2016).
  doi:10.1142/S0217732316300147
  %%CITATION = doi:10.1142/S0217732316300147;%%



%\cite{Antusch:2005kf}
\bibitem{Antusch:2005kf} 
  S.~Antusch, O.~J.~Eyton-Williams and S.~F.~King,
  %``Dirac neutrinos and hybrid inflation from string theory,''
  JHEP {\bf 0508}, 103 (2005)
  doi:10.1088/1126-6708/2005/08/103
  [hep-ph/0505140].
  %%CITATION = doi:10.1088/1126-6708/2005/08/103;%%
  %20 citations counted in INSPIRE as of 05 Feb 2020



%\cite{Langacker:2011bi}
\bibitem{Langacker:2011bi} 
  P.~Langacker,
  %``Neutrino Masses from the Top Down,''
  Ann.\ Rev.\ Nucl.\ Part.\ Sci.\  {\bf 62}, 215 (2012)
  doi:10.1146/annurev-nucl-102711-094925
  [arXiv:1112.5992 [hep-ph]].
  %%CITATION = doi:10.1146/annurev-nucl-102711-094925;%%
  %24 citations counted in INSPIRE as of 05 Feb 2020



%\cite{Dvali:2016uhn}
\bibitem{Dvali:2016uhn} 
  G.~Dvali and L.~Funcke,
  %``Small neutrino masses from gravitational θ-term,''
  Phys.\ Rev.\ D {\bf 93}, no. 11, 113002 (2016)
  doi:10.1103/PhysRevD.93.113002
  [arXiv:1602.03191 [hep-ph]].
  %%CITATION = doi:10.1103/PhysRevD.93.113002;%%
  %46 citations counted in INSPIRE as of 05 Feb 2020



%\cite{Barenboim:2019fmj}
\bibitem{Barenboim:2019fmj} 
  G.~Barenboim, J.~Turner and Y.~L.~Zhou,
  %``Light Neutrino Masses from Gravitational Condensation: the Schwinger-Dyson Approach,''
  arXiv:1909.04675 [hep-ph].
  %%CITATION = ARXIV:1909.04675;%%
  %1 citations counted in INSPIRE as of 05 Feb 2020



%\cite{Broggini:2012df}
\bibitem{Broggini:2012df} 
  C.~Broggini, C.~Giunti and A.~Studenikin,
  %``Electromagnetic Properties of Neutrinos,''
  Adv.\ High Energy Phys.\  {\bf 2012}, 459526 (2012)
  doi:10.1155/2012/459526
  [arXiv:1207.3980 [hep-ph]].
  %%CITATION = doi:10.1155/2012/459526;%%
  %89 citations counted in INSPIRE as of 05 Feb 2020



%\cite{Kobayashi:1973fv}
\bibitem{Kobayashi:1973fv} 
  M.~Kobayashi and T.~Maskawa,
  %``CP Violation in the Renormalizable Theory of Weak Interaction,''
  Prog.\ Theor.\ Phys.\  {\bf 49}, 652 (1973).
  doi:10.1143/PTP.49.652
  %%CITATION = doi:10.1143/PTP.49.652;%%
  %10130 citations counted in INSPIRE as of 05 Feb 2020



%\cite{Maiani:1975in}
\bibitem{Maiani:1975in} 
  L.~Maiani,
  %``CP Violation in Purely Lefthanded Weak Interactions,''
  Phys.\ Lett.\  {\bf 62B}, 183 (1976).
  doi:10.1016/0370-2693(76)90500-1
  %%CITATION = doi:10.1016/0370-2693(76)90500-1;%%
  %333 citations counted in INSPIRE as of 05 Feb 2020



%\cite{Branco:1988iq}
\bibitem{Branco:1988iq} 
  G.~C.~Branco, L.~Lavoura and F.~Mota,
  %``Nearest Neighbor Interactions and the Physical Content of Fritzsch Mass Matrices,''
  Phys.\ Rev.\ D {\bf 39}, 3443 (1989).
  doi:10.1103/PhysRevD.39.3443
  %%CITATION = doi:10.1103/PhysRevD.39.3443;%%
  %107 citations counted in INSPIRE as of 05 Feb 2020



%\cite{Branco:1999nb}
\bibitem{Branco:1999nb} 
  G.~C.~Branco, D.~Emmanuel-Costa and R.~Gonzalez Felipe,
  %``Texture zeros and weak basis transformations,''
  Phys.\ Lett.\ B {\bf 477}, 147 (2000)
  doi:10.1016/S0370-2693(00)00193-3
  [hep-ph/9911418].
  %%CITATION = doi:10.1016/S0370-2693(00)00193-3;%%
  %94 citations counted in INSPIRE as of 05 Feb 2020



%\cite{Tanabashi:2018oca}
\bibitem{Tanabashi:2018oca} 
  M.~Tanabashi {\it et al.} [Particle Data Group],
  %``Review of Particle Physics,''
  Phys.\ Rev.\ D {\bf 98}, no. 3, 030001 (2018).
  doi:10.1103/PhysRevD.98.030001
  %%CITATION = doi:10.1103/PhysRevD.98.030001;%%
  %3725 citations counted in INSPIRE as of 05 Feb 2020



%\cite{Fusaoka:1998vc}
\bibitem{Fusaoka:1998vc} 
  H.~Fusaoka and Y.~Koide,
  %``Updated estimate of running quark masses,''
  Phys.\ Rev.\ D {\bf 57}, 3986 (1998)
  doi:10.1103/PhysRevD.57.3986
  [hep-ph/9712201].
  %%CITATION = doi:10.1103/PhysRevD.57.3986;%%
  %336 citations counted in INSPIRE as of 05 Feb 2020



%\cite{Giraldo:2011ya}
\bibitem{Giraldo:2011ya} 
  Y.~Giraldo,
  %``Texture Zeros and WB Transformations in the Quark Sector of the Standard Model,''
  Phys.\ Rev.\ D {\bf 86}, 093021 (2012)
  doi:10.1103/PhysRevD.86.093021
  [arXiv:1110.5986 [hep-ph]].
  %%CITATION = doi:10.1103/PhysRevD.86.093021;%%
  %13 citations counted in INSPIRE as of 05 Feb 2020



%\cite{Giraldo:2015cpp}
\bibitem{Giraldo:2015cpp} 
  Y.~Giraldo,
  %``Five-Zero Texture Non-Fritzsch like Quark Mass Matrices in the Standard Model,''
  arXiv:1511.08858 [hep-ph].
  %%CITATION = ARXIV:1511.08858;%%
  %4 citations counted in INSPIRE as of 05 Feb 2020



%\cite{Verma:2016qhy}
\bibitem{Verma:2016qhy} 
  R.~Verma,
  %``Implications of $\delta^{CP}_l\sim 270^\circ$ and $\theta_{23}\gtrsim 45^\circ$ for texture specific lepton mass matrices and $0\nu \beta \beta$ decay,''
  Adv.\ High Energy Phys.\  {\bf 2016}, 2094323 (2016)
  doi:10.1155/2016/2094323
  [arXiv:1607.00958 [hep-ph]].
  %%CITATION = doi:10.1155/2016/2094323;%%
  %3 citations counted in INSPIRE as of 05 Feb 2020



%\cite{Esteban:2018azc}
\bibitem{Esteban:2018azc} 
  I.~Esteban, M.~C.~Gonzalez-Garcia, A.~Hernandez-Cabezudo, M.~Maltoni and T.~Schwetz,
  %``Global analysis of three-flavour neutrino oscillations: synergies and tensions in the determination of $\theta_{23}$, $\delta_{CP}$, and the mass ordering,''
  JHEP {\bf 1901}, 106 (2019)
  doi:10.1007/JHEP01(2019)106
  [arXiv:1811.05487 [hep-ph]].
  %%CITATION = doi:10.1007/JHEP01(2019)106;%%
  %274 citations counted in INSPIRE as of 05 Feb 2020



%\cite{Rasin:1997pn}
\bibitem{Rasin:1997pn} 
  A.~Rasin,
  %``Diagonalization of quark mass matrices and the Cabibbo-Kobayashi-Maskawa matrix,''
  hep-ph/9708216.
  %%CITATION = HEP-PH/9708216;%%
  %37 citations counted in INSPIRE as of 05 Feb 2020

\end{thebibliography}
%\acknowledgements 
%\input{references.tex}

\end{document}